\documentclass[journal]{IEEEtran}

\usepackage{cite}

\newcommand\beq{\begin{equation}}
\newcommand\eeq{\end{equation}}

\usepackage{graphicx}
\usepackage{dcolumn}
\usepackage{bm}

\usepackage[utf8x]{inputenc}

\usepackage{amsmath}
\usepackage{amssymb}

\usepackage{array}

\usepackage{stfloats}
\begin{document}
\title{Surface-Wave Propagation on Non-Hermitian Metasurfaces with Extreme Anisotropy}
%
%
%

\author{Marino~Coppolaro,
	Massimo~Moccia,
        Giuseppe~Castaldi,
        Andrea~Al\`u,
        and  Vincenzo Galdi
\thanks{M. Coppolaro, M. Moccia, G. Castaldi, and V. Galdi are with the Fields \& Waves Lab, Department of Engineering, University of Sannio, I-82100 Benevento, Italy (e-mail: coppolaro@unisannio.it, massimo.moccia@unisannio.it, castaldi@unisannio.it, vgaldi@unisannio.it).}
\thanks{A. Al\`u is with the Photonics Initiative, Advanced
	Science Research Center, the Physics Program, Graduate Center, and the
	Department of Electrical Engineering, City College, all at the City University
	of New York, New York, NY 10031, USA (email: aalu@gc.cuny.edu)}
}

\date{\today}


%



\maketitle

\begin{abstract}
Electromagnetic metasurfaces enable the advanced control of surface-wave propagation by spatially tailoring the local surface reactance. 
Interestingly, tailoring the surface resistance distribution in space provides new, largely unexplored degrees of freedom. Here, we show that suitable spatial modulations of the surface resistance between positive (i.e., loss) and negative (i.e., gain) values can induce peculiar dispersion effects, far beyond a mere compensation. Taking inspiration from the parity-time symmetry concept in quantum physics, we put forward and explore a class of non-Hermitian metasurfaces that may exhibit extreme anisotropy mainly induced by the gain-loss interplay. Via analytical modeling and full-wave numerical simulations, we illustrate the associated phenomenon of surface-wave canalization, explore nonlocal effects and possible departures from the ideal conditions, and address the feasibility of the required constitutive parameters.
Our results suggest intriguing possibilities to dynamically reconfigure the surface-wave propagation, and are of potential interest for applications to imaging, sensing and communications.
\end{abstract}


%
\IEEEpeerreviewmaketitle

\section{Introduction}

\IEEEPARstart
{S}{urface electromagnetics} is a research topic of longstanding interest in microwave and antenna engineering, which is experiencing a renewed vitality (see, e.g., \cite{Yang:2019se} for a recent review) in view of the widespread applications of artificial  (metasurfaces) \cite{Holloway:2012ao}
 and natural (e.g., graphene) \cite{Bao:20192d} low-dimensional (2-D) materials.

In addition to enabling advanced wavefront manipulations \cite{Yu:2011lp}, metasurfaces can support the propagation of tightly bound surface waves \cite{Bilow:2003gw,Patel:2013ma,Quarfoth:2013at,Mencagli:2015sw}
which can be finely controlled via transformation-optics approaches \cite{Vakil:2011to,Yang:2012aa,Patel:2014te,Mencagli:2014mb,Martini:2015mt,McCall:2018ro}
conceptually similar to those applied for volumetric metamaterials \cite{Pendry:2006e,Leonhardt:2006oc}. Likewise, exotic phenomena observed in volumetric metamaterials can be transposed  to ``flatland'' scenarios. These include, for instance, hyperbolic propagation (characterized by open dispersion characteristics) 
\cite{Yermakov:2015hw,Gomez-Diaz:2015hp,Gomez-Diaz:2015hy,Gomez-Diaz:2016fo,Yang:2017hs,Yermakov:2018es}, 
topological transitions (from closed elliptic-like to open hyperbolic-like dispersion characteristics) \cite{Gomez-Diaz:2015hp}, 
extreme anisotropy (i.e., very elongated dispersion characteristics) \cite{Gomez-Diaz:2017pc}, and canalization (i.e., diffractionless propagation of subwavelength beams) \cite{Gomez-Diaz:2015hp,Gomez-Diaz:2015hy,Gomez-Diaz:2016fo,Gomez-Diaz:2017pc}.
Interestingly, new intriguing concepts and effects are emerging that are specific of 2-D materials. Among these, it is worth mentioning the ``line waves'', localized both
in-plane and out-of-plane around a surface reactance discontinuity with
dual character (capacitive/inductive) \cite{Horsley:2014od,Bisharat:2017ge}, and the rich
moir\'e physics observed in rotated, evanescently coupled metasurfaces
\cite{Hu:2020mh,Hu:2020tp}.

The above studies have focused on {\em passive} scenarios, wherein unavoidable losses are undesired and their effects need to be minimized. In fact, it has been shown that, in certain parameter regimes, losses can be beneficially exploited to enhance the canalization effects \cite{Gomez-Diaz:2017pc}. In this study, we further leverage and generalize this concept by exploring a class of metasurfaces characterized by tailored spatial modulations of loss and gain. Inspired by quantum-physics concepts such as ``parity-time'' (${\cal PT}$) symmetry
\cite{Bender:1998rs}, these non-Hermitian configurations are garnering a growing interest in several branches of physics \cite{El-Ganainy:2018nh}, including electromagnetics \cite{Feng:2017nh}. In quantum mechanics, a ${\cal PT}$-symmetric operator is characterized by a potential function that satisfies the condition $V\left(-x\right)=V^*\left(x\right)$, with $x$ and $^*$ denoting a spatial coordinate and complex conjugation, respectively \cite{Bender:1998rs}. In view of the well-known analogies, in electromagnetic scenarios, this translates into a refractive-index distribution $n\left(-x\right)=n^*\left(x\right)$, which implies an imaginary part with {\em odd} symmetry, corresponding to alternated loss and gain regions. Likewise, when referred to metasurfaces, of specific interest in our study, this condition translates as $Z\left(-x\right)=-Z^*\left(x\right)$ and $\sigma\left(-x\right)=-\sigma^*\left(x\right)$ for the surface impedance and conductivity, respectively.
Within this framework, several non-Hermitian metasurface scenarios have been explored, with possible applications to
negative refraction \cite{Fleury:2014nr},
cloaking \cite{Sounas:2015uc},
imaging \cite{Monticone:2016pt,Savoia:2017mi},
sensing \cite{Chen:2016PT,Sakhdari:2017pt,Farhat:2020},
low-threshold laser and coherent perfect absorbers \cite{Sakhdari:2018},
line waves \cite{Moccia:2020lw}, and
unconventional lattice resonances \cite{Kolkowski:2020lr}. However, in most studies gain and loss are distributed on separate metasurfaces, with {\em out-of-plane} coupling, and there are only few examples of metasurfaces featuring {\em in-plane} modulation of gain and loss. In fact, the effect of the gain-loss interplay in the surface-wave propagation remains largely unexplored.

Here, inspired by recent findings for volumetric metamaterials \cite{Coppolaro:2020ep}, we show that the judicious tailoring
of the gain-loss interplay can induce {\em extreme-anisotropy} responses in
non-Hermitian metasurfaces. The anisotropy is particularly pronounced under ${\cal PT}$-symmetry conditions, and yields strong surface-wave canalization effects that significantly depend on the gain-loss level. The possibility to control the gain (e.g., via solid-state amplifiers or optical pumping, depending on the operational frequency) opens the door to intriguing strategies for dynamic reconfiguration of the response. 

The rest of the paper is structured as follows. In Sec. \ref{Sec:Math}, we outline the problem formulation and its geometry.
In Sec. \ref{Sec:MR}, we derive the parameter regimes of interest and illustrate some representative results, obtained via an effective-medium theory (EMT) and full-wave simulations, with details relegated in two Appendices. Specifically, we study the extreme anisotropy and associated canalization phenomena that can occur under ${\cal PT}$-symmetry conditions, as well as the effects of nonlocality and possible departures from these conditions.
In Sec. \ref{Sec:Implementation}, we explore the practical feasibility of the required gain levels, also addressing the stability issues. 
Finally, in Sec. \ref{Sec:Conclusions}, we draw some brief conclusions and discuss possible perspectives.

%
\begin{figure}
	\centering
	\includegraphics[width=.8\linewidth]{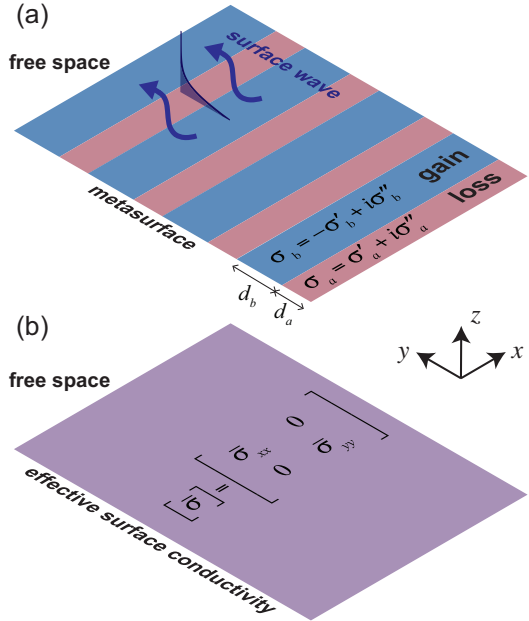}
	\caption{(a) Problem geometry (details in the text). (b) Effective-parameter representation.}
	\label{Figure1}
\end{figure}

\section{Problem Geometry and Formulation}
\label{Sec:Math}

\subsection{Geometry}
The problem geometry is schematically illustrated in Fig. \ref{Figure1}a. We consider, in free space, a metasurface of infinite extent in the $x-y$ plane, featuring a 1-D periodic modulation of the surface conductivity along the $y$-direction, with alternating values $\sigma_a$ and $\sigma_b$ (and sub-periods $d_a$ and $d_b$, respectively). Assuming an implicit $\exp(-i\omega t)$ time-harmonic  dependence, the
generally complex-valued conductivities are written as
\beq
{\sigma _a} = {\sigma^\prime_a} + i{\sigma^{\prime\prime}_a},\quad {\sigma _b} =  - {\sigma^\prime_b} + i{\sigma^{\prime\prime}_b},
\label{eq:sigmaab}
\eeq
with the prime and double-prime symbols tagging the real and imaginary parts, respectively. Throughout the study, we focus on the parameter regime
\beq
 \sigma^\prime_{a,b} > 0,\quad
 \sigma^{\prime\prime}_a \sigma^{\prime\prime}_b> 0.
 \label{eq:ass}
 \eeq 
 In view of the assumed time-dependence, the first condition implies that the ``$a$''- and ``$b$''-type constituents exhibit loss and gain, respectively. The second condition instead guarantees that both constituents are either of {\em inductive} ($\sigma^{\prime\prime}_{a,b}>0$) or {\em capacitive} ($\sigma^{\prime\prime}_{a,b}<0$) nature; this rules out the possibility of a {\em hyperbolic} response, which is not of interest here since it has already been studied \cite{Gomez-Diaz:2015hy,Gomez-Diaz:2016fo}.
 
We highlight that the surface-conductivity parameterization in (\ref{eq:sigmaab}) and (\ref{eq:ass}) is especially suited for 2-D materials (e.g., graphene), and can be readily related to more conventional constitutive parameters. For instance, it can be obtained from the surface impedance as \cite{Bisharat:2018ml}
  \beq
 \sigma=\frac{2}{Z},
 \eeq
 with the factor 2 accounting for the two-faced character of the sheet. Moreover, for a very thin dielectric layer of thickness $\Delta\ll \lambda$  (with $\lambda=2\pi c/\omega$ denoting the free-space wavelength and $c$ the corresponding wavespeed), it can be approximately related to the relative permittivity via \cite{Vakil:2011to,Mattheakis:2016en}
\beq
\sigma\approx
\frac{i\left( {1 - \varepsilon } \right)k\Delta}{\eta},
\label{eq:sigd}
\eeq 
where $k=\omega/c=2\pi/\lambda$ and $\eta=\sqrt{\mu_0/\varepsilon_0}$ denote the free-space wavenumber and characteristic impedance, respectively. 
 
\subsection{Formulation}
In \cite{Forati:2014ph}, it was shown that, in the limit of {\em deeply subwavelength}  modulation periods $d=d_a+d_b\ll \lambda$, a structure like the one in Fig. \ref{Figure1}a could be effectively modeled by a homogeneous, uniaxially anisotropic effective surface conductivity with relevant components
\beq
{\bar \sigma _{xx}} = {f_a}{\sigma _a} + {f_b}{\sigma _b},\quad {\bar \sigma _{yy}} = \frac{{{\sigma _a}{\sigma _b}}}{{{f_b}{\sigma _a} + {f_a}{\sigma _b}}},
\label{eq:EMT}
\eeq
with $f_a=d_a/d$   and $f_b=1-f_a$   denoting the filling fractions. These mixing formulae closely resemble those occurring in the EMT modeling of multilayered volumetric metamaterials \cite{Sihvola:1999em}.

We are interested in studying the propagation of surface waves along this metasurface. In mathematical terms, this entails finding nontrivial source-free solutions $\propto \exp\left[i\left(k_x x+k_y y+k_z z\right)\right]$ which are evanescent along the $z$-direction and propagating in the $x-y$ plane (see Fig. \ref{Figure1}). By assuming for now the EMT model in (\ref{eq:EMT}), it can be shown (see \cite{Bilow:2003gw,Patel:2013ma,Quarfoth:2013at} for details) that the wavenumbers must satisfy the dispersion equation
\begin{subequations}
	\begin{eqnarray}
&&k{k_z}\left( {4 + {\eta ^2}{{\bar \sigma }_{xx}}{{\bar \sigma }_{yy}}} \right) \nonumber\\
	&+& 2\eta \left[ {{k^2}\left( {{{\bar \sigma }_{xx}} + {{\bar \sigma }_{yy}}} \right) - {{\bar \sigma }_{xx}}k_x^2 - {{\bar \sigma }_{yy}}k_y^2} \right] = 0,
	\label{eq:DE}
	\end{eqnarray}
with the constraints
	\beq
	k_x^2 + k_y^2 + k_z^2 = {k^2},\quad {\mathop{\rm Im}\nolimits} \left( {{k_z}} \right) \ge 0.
	\label{eq:PWD}
	\eeq
	\label{eq:DEs}
\end{subequations}
The numerical solution of the nonlinear system in (\ref{eq:DEs}) can be efficiently carried out following the approach described in \cite{Gomez-Diaz:2015hy}.
The resulting modes are generally hybrid, and contain as special cases the transverse-electric and transverse-magnetic modes that can be supported by isotropic, inductive and capacitive metasurfaces, respectively.
Moreover, it is readily verified that the dispersion equation in (\ref{eq:DE}) is invariant under the duality transformations
\beq
{\bar \sigma _{xx}} \to \frac{4}{{{\eta ^2}{{\bar \sigma }_{yy}}}},\quad {\bar \sigma _{yy}} \to \frac{4}{{{\eta ^2}{{\bar \sigma }_{xx}}}}.
\label{eq:duality}
\eeq
The above relationships somehow resemble those observed for self-complementary metasurfaces \cite{Ortiz:2013sc,Gonzalez:2015sw,Baena:2017ba}, which exploit the Babinet's principle. In our case, rather than geometrical self-complementarity, we rely on ${\cal PT}$ symmetry.

In what follows, we will study the effects of the gain-loss interplay in the surface-wave propagation, via the approximate EMT modeling and full-wave numerical simulations.

\section{Modeling and Results}
\label{Sec:MR}
\subsection{Loss-Gain Compensation in Effective Parameters}
Looking at the mixing formulae in (\ref{eq:EMT}), an interesting question is whether there exist specific combinations of the constituents (in terms of $\sigma_a$, $\sigma_b$ and $f_a$) that render the effective parameters {\em purely reactive}. Within the limitations of the EMT, this would imply a perfect balancing of the loss and gain effects, which in the language of ${\cal PT}$-symmetry, corresponds to the symmetric phase \cite{Bender:1998rs}.
By substituting the complex-valued conductivities (\ref{eq:sigmaab}) in the mixing formulae (\ref{eq:EMT}), and enforcing the zeroing of the effective-parameter real parts, we obtain the conditions (see Appendix \ref{Sec:AppA} for details)
\begin{subequations}
\begin{eqnarray}
f_a &=& \frac{\sigma^{\prime}_b}{\sigma^{\prime}_a+\sigma^{\prime}_b},
\label{eq:fa}\\
{\sigma^{\prime\prime}_a} &=&  \pm 
\sqrt{\left|\sigma_b\right|^2-\left(\sigma^{\prime}_a\right)^2}.
\label{eq:sigmappa}
\end{eqnarray}
\label{eq:lcEP}
\end{subequations}
In view of the assumption $\sigma^{\prime}_{a,b}>0$, the condition in (\ref{eq:fa}) is always feasible ($0\le f_a\le1$). On the other hand, we must enforce the additional constraint
\beq
\left|\sigma_b\right|>\sigma^{\prime}_a,
\eeq
in order to ensure that ${\sigma^{\prime\prime}_a}$ in (\ref{eq:sigmappa}) is consistently real-valued. Moreover, in view of our assumption $\sigma^{\prime\prime}_a \sigma^{\prime\prime}_b> 0$, the sign determination in (\ref{eq:sigmappa}) must be chosen consistently with the sign of $\sigma^{\prime\prime}_b$.
By enforcing the conditions (\ref{eq:lcEP}) in (\ref{eq:EMT}), after some algebra, we obtain 
\begin{subequations}
\begin{eqnarray}
{\bar \sigma}_{xx}&=&i\frac{\sigma^{\prime}_a\sigma^{\prime\prime}_b\pm\sigma^{\prime}_b
\sqrt{\left|\sigma_b\right|^2-\left(\sigma^{\prime}_a\right)^2}}
{\sigma^{\prime}_a+\sigma^{\prime}_b},\\
{\bar \sigma}_{yy}&=&i\frac{\sigma^{\prime}_a\sigma^{\prime\prime}_b\mp \sigma^{\prime}_b
	\sqrt{\left|\sigma_b\right|^2-\left(\sigma^{\prime}_a\right)^2}}
{\sigma^{\prime}_a-\sigma^{\prime}_b},
\label{eq:EPb}
\end{eqnarray}
\label{eq:EP}
\end{subequations}
which identify a class of  parameter combinations that yield {\em purely imaginary} effective parameters, i.e., loss-gain balance.  
This result is not necessarily surprising, as a balanced amount of gain and loss may be expected to compensate each other,
but we will show hereafter that the gain-loss interplay may have much deeper implications on the modes supported by this metasurface.

%
\begin{figure}
	\centering
	\includegraphics[width=.8\linewidth]{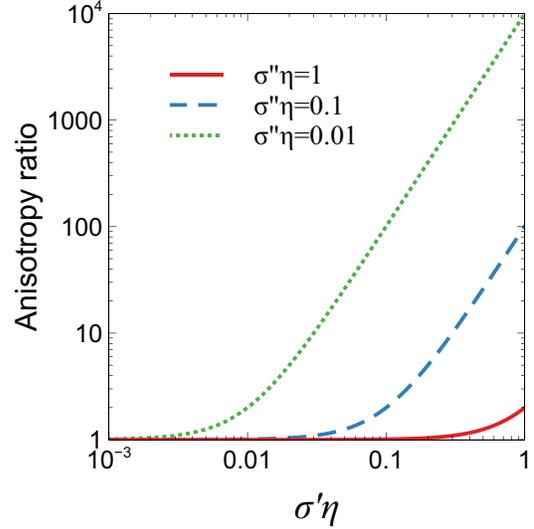}
	\caption{Anisotropy ratio $\left|{\bar\sigma}_{yy}\right|/\left|{\bar\sigma}_{xx}\right|$ for ${\cal PT}$-symmetric configurations, as a function of the gain-loss parameter $\sigma^\prime\eta$, for representative values of the normalized susceptance: $\sigma^{\prime\prime}\eta=1$ (red-solid curve),  $\sigma^{\prime\prime}\eta=0.1$ (blue-solid), and $\sigma^{\prime\prime}\eta=0.01$ (green-dotted). Note the log-log scale.}
	\label{Figure2}
\end{figure}

%
\begin{figure*}
	\centering
	\includegraphics[width=\linewidth]{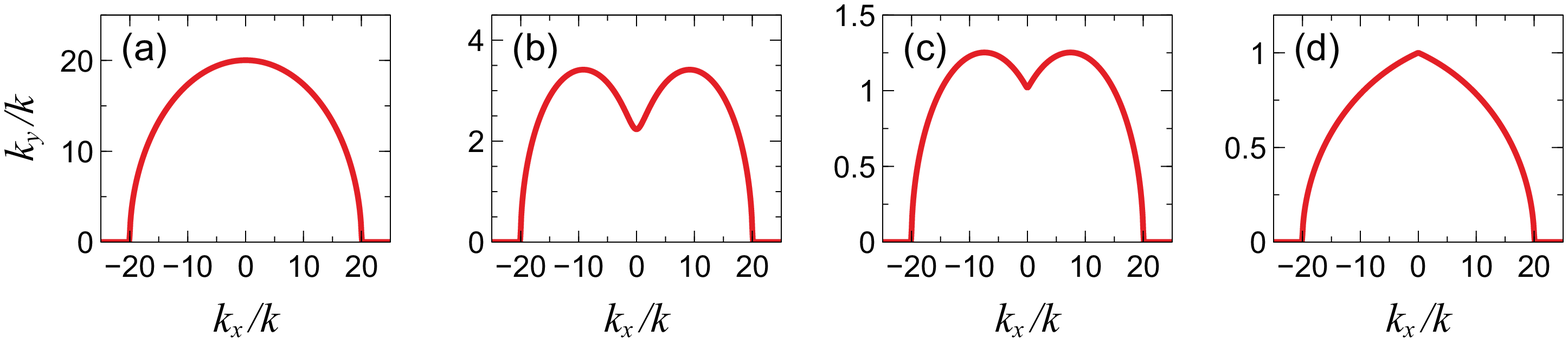}
	\caption{Examples of EFCs for ${\cal PT}$-symmetric inductive configurations with $\sigma^{\prime\prime}\eta=0.1$. 
		(a) $\sigma^{\prime}=0$ (${\bar \sigma}_{xx} \eta={\bar \sigma}_{yy} \eta=i0.1$, $\left|{\bar \sigma}_{yy} \right|/\left|{\bar \sigma}_{xx} \right|=1$).
		(b) $\sigma^{\prime}\eta=0.3$ (${\bar \sigma}_{xx} \eta=i0.1$, ${\bar \sigma}_{yy} \eta=i$, $\left|{\bar \sigma}_{yy} \right|/\left|{\bar \sigma}_{xx} \right|=10$).
		(c) $\sigma^{\prime}\eta=1$ (${\bar \sigma}_{xx} \eta=i0.1$, ${\bar \sigma}_{yy} \eta=i10.1$, $\left|{\bar \sigma}_{yy} \right|/\left|{\bar \sigma}_{xx} \right|=101$). 
		(d) $\sigma^{\prime}\eta=3$, (${\bar \sigma}_{xx} \eta=i0.1$, ${\bar \sigma}_{yy} \eta=i90.1$, $\left|{\bar \sigma}_{yy} \right|/\left|{\bar \sigma}_{xx} \right|=901$).
		In view of the inherent symmetry, only the $k_y>0$ branches are shown. Note the different scales on the vertical axes.}
	\label{Figure3}
\end{figure*}

%
\begin{figure*}
	\centering
	\includegraphics[width=\linewidth]{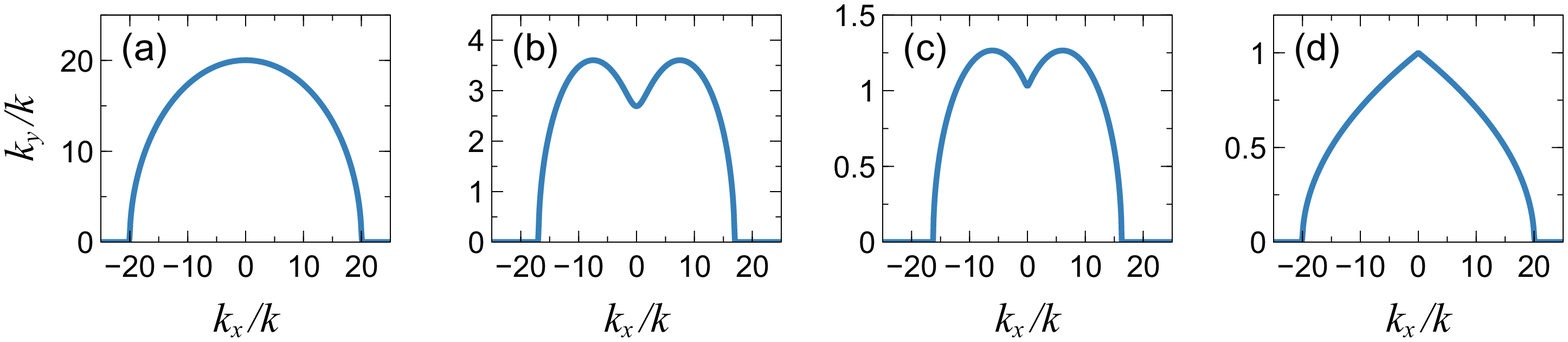}
	\caption{Examples of EFCs for ${\cal PT}$-symmetric capacitive configurations. (a) 
		$\sigma^{\prime}=0$ and $\sigma^{\prime\prime}\eta=-40$ (${\bar \sigma}_{xx} \eta={\bar \sigma}_{yy} \eta=-i40$, $\left|{\bar \sigma}_{yy} \right|/\left|{\bar \sigma}_{xx} \right|=1$).
		(b) $\sigma^{\prime}\eta=12$ and $\sigma^{\prime\prime}\eta=-5$ (${\bar \sigma}_{xx} \eta=-i5$, ${\bar \sigma}_{yy} \eta=-i33.8$, $\left|{\bar \sigma}_{yy} \right|/\left|{\bar \sigma}_{xx} \right|=6.76$).
		(c) $\sigma^{\prime}\eta=4$ and $\sigma^{\prime\prime}\eta=-0.5$ (${\bar \sigma}_{xx} \eta=-i0.5$, ${\bar \sigma}_{yy} \eta=-i32.5$, $\left|{\bar \sigma}_{yy} \right|/\left|{\bar \sigma}_{xx} \right|=65$).
		(d) $\sigma^{\prime}\eta=0.2$ and $\sigma^{\prime\prime}\eta=-0.001$ (${\bar \sigma}_{xx} \eta=-i0.001$, ${\bar \sigma}_{yy} \eta=-i40$, $\left|{\bar \sigma}_{yy} \right|/\left|{\bar \sigma}_{xx} \right|=4000$). 
		In view of the inherent symmetry, only the $k_y>0$ branches are shown. Note the different scales on the vertical axes.}
	\label{Figure4}
\end{figure*}

\subsection{${\cal PT}$-Symmetry-Induced Extreme Anisotropy}
By inspection of (\ref{eq:EP}), it can be observed that the conductivity components can differ substantially in the limit
$\sigma^{\prime}_a\rightarrow \sigma^{\prime}_b$. From (\ref{eq:lcEP}), this yields $f_a=f_b=0.5$ and (with the sign determination of interest here) ${\sigma^{\prime\prime}_a}={\sigma^{\prime\prime}_b}$. In other words, by removing irrelevant superscripts, we obtain
\beq
\sigma_a=\sigma^\prime+i \sigma^{\prime\prime},\quad
\sigma_b=-\sigma^\prime+i \sigma^{\prime\prime},
\label{eq:PTSab}
\eeq
which, with a suitable choice of the reference-system origin, corresponds to the aforementioned ${\cal PT}$-symmetry condition
\beq
\sigma\left(-y\right)=-\sigma^*\left(y\right).
\label{eq:PTS}
\eeq
It can be shown (see Appendix \ref{Sec:AppA} for details) that, under these conditions, the effective parameters reduce to
\beq
{\bar \sigma}_{xx}=i\sigma^{\prime\prime},\quad
{\bar \sigma}_{yy}=i\frac{\left|\sigma\right|^2}{\sigma^{\prime\prime}}.
\label{eq:EP1}
\eeq
The remarkably simple expressions in (\ref{eq:EP1}) clearly show that the gain-loss interplay can significantly affect the effective-parameter anisotropy. 
In particular, it is evident that the conditions
\beq
\left|\sigma^{\prime\prime}\right|\eta\ll 1, \quad \sigma^{\prime}\gg\left|\sigma^{\prime\prime}\right|
\label{eq:CAE}
\eeq
would lead to $\left|{\bar \sigma}_{xx}\right|\ll\left|{\bar \sigma}_{yy}\right|$, i.e., {\em extreme anisotropy}. For a quantitative illustration, Fig. \ref{Figure2} shows the anisotropy ratio $\left|{\bar \sigma}_{yy}\right|/\left|{\bar \sigma}_{xx}\right|$ as a function of the gain-loss parameter ${\bar \sigma}^{\prime}\eta$ for various values of the normalized susceptance $\sigma^{\prime\prime}\eta$; note that results do not depend on the sign of $\sigma^{\prime\prime}$, i.e., on the inductive/capacitive character. However, for a given anisotropy ratio, the dispersion characteristics do depend on the reactive character of the metasurface. 

For example, Fig. \ref{Figure3} shows some representative dispersion characteristics, in terms of equi-frequency contours (EFCs), 
for an inductive ($\bar \sigma^{\prime\prime} > 0$) scenario, by fixing $\sigma^{\prime\prime}\eta=0.1$ and varying  the gain-loss parameter $\sigma^{\prime}\eta$. As can be observed, the field exhibits a propagating character within a spectral wavenumber region that can be estimated analytically from (\ref{eq:DEs}) (see Appendix \ref{Sec:AppA} for details), viz.
\beq
\left|k_x\right|\le k_x^{\left( {\max } \right)} =  k
\sqrt{1-\frac{4}{{\bar \sigma _{xx}^2{\eta ^2}}}}
	\approx  
 \frac{{2k}}{{\sigma^{\prime\prime}}\eta},
 \label{eq:kx1}
\eeq
with the approximate equality holding in the limit $\sigma^{\prime\prime}\eta  \ll 1$. Quite interestingly, this spatial bandwidth is essentially controlled by the normalized susceptance $\sigma^{\prime\prime}\eta$. Outside this region, the field is evanescent, with a purely imaginary $k_y$ (not shown for brevity). The gain-loss parameter $\sigma^{\prime}\eta$ controls instead the anisotropy degree. Specifically, starting from the trivial ($\sigma^\prime=0$) isotropic case (Fig. \ref{Figure3}a), the anisotropy becomes increasingly pronounced by increasing $\sigma^\prime$ (Figs. \ref{Figure3}b and \ref{Figure3}c), and the EFCs approach a limiting curve for $\sigma^\prime\eta\gg1$ (Fig. \ref{Figure3}d). 

The responses for capacitive ($\bar \sigma^{\prime\prime} > 0$) scenarios can be in principle obtained via the duality transformations in (\ref{eq:duality}). However, for completeness, they are also exemplified in Fig. \ref{Figure4} directly in terms of the constituent parameters $\sigma^{\prime}$ and $\sigma^{\prime\prime}$. In this case, the propagating spectral region is given by (see Appendix \ref{Sec:AppA} for details)
\beq
\left|k_x\right|\le k_x^{\left( {\max } \right)} =  k \sqrt {1 - \frac{\bar \sigma _{yy}^2{\eta ^2}}{4}}  \approx
\frac{{k\eta\left(\sigma^\prime\right)^2}}{2\left|\sigma^{\prime\prime}\right|},
\label{eq:kx2}
\eeq
where the approximate equality holds in the asymptotic limit $\eta\left(\sigma^\prime\right)^2 \gg \left|\sigma^{\prime\prime}\right|$. We observe that, unlike the inductive case, the spatial bandwidth now depends on both $\sigma^{\prime}$ and $\sigma^{\prime\prime}$; the representative values considered in Fig. \ref{Figure4}  are chosen so as to maintain the spatial bandwidth $k_x^{(max)}\approx 20 k$, progressively moving from perfect isotropy (Fig. \ref{Figure4}a) to extreme anisotropy (Fig. \ref{Figure4}d).

%
\begin{figure*}
	\centering
	\includegraphics[width=.8\linewidth]{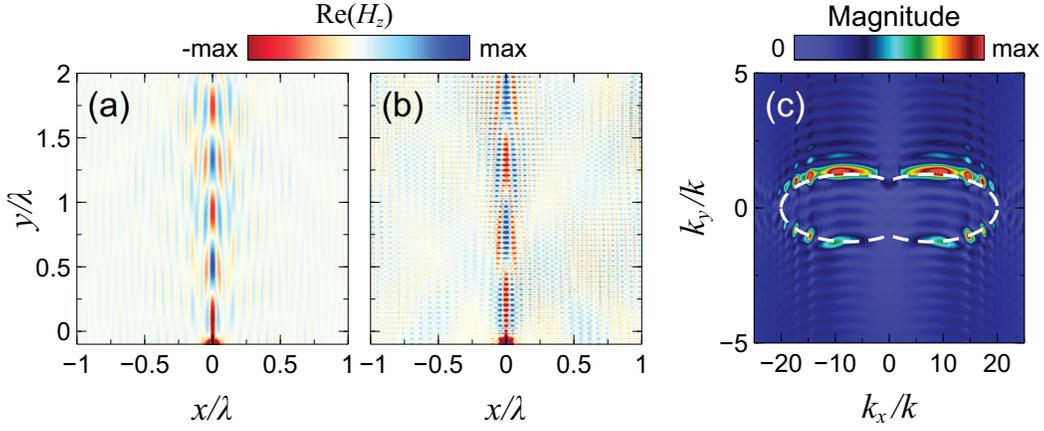}
	\caption{Examples of canalization effects. (a) Numerically computed in-plane field map [$\mbox{Re}\left(H_z\right)$], in false-color scale, pertaining to a metasurface with effective parameters ${\bar \sigma_{xx}}\eta=i0.1$, ${\bar \sigma_{yy}}\eta=i10.1$. (b) Corresponding results for the actual ${\cal PT}$-symmetric conductivity modulation, with $\sigma_a\eta=1+i0.1$, $\sigma_b\eta=-1+i0.1$, $f_a=f_b=0.5$, and $d=0.025\lambda$. Fields are excited by a $z$-directed elementary magnetic dipole located at $x=0$, $y=-0.1\lambda$, $z=0.001\lambda$, and are computed at $z=0.01\lambda$. (c) Spatial spectrum (2-D Fourier transform) magnitude, in false color scale, of the field map in panel (a). Also shown (white dashed curve), as a reference, is the theoretical EFC from Fig. \ref{Figure3}c.}
	\label{Figure5}
\end{figure*}

\subsection{Canalization Effects}
Canalization effects, intended as the diffractionless transfer of subwavelength features over distances of several wavelengths, have been observed in hyperbolic \cite{Gomez-Diaz:2015hp,Gomez-Diaz:2015hy} and extreme-anisotropy \cite{Gomez-Diaz:2017pc} metasurfaces, and loss-induced canalization has also been demonstrated \cite{Jiang:2015ev}.

These effects can be intuitively understood by looking at the examples in Figs. \ref{Figure3} and \ref{Figure4} with higher anisotropy (e.g., Figs. \ref{Figure3}c, \ref{Figure3}d, \ref{Figure4}c and \ref{Figure4}d). It is apparent that a significant fraction of high-$k_x$  spectral components can propagate as unattenuated surface waves in the $x-y$ plane and, in view of the pronounced flatness of the EFCs, their group velocities (normal to the EFCs) are predominantly $y$-directed.

For illustration, Fig. \ref{Figure5} shows the surface-wave propagation along an inductive, non-Hermitian metasurface (with parameters as in Fig. \ref{Figure3}c), excited by a $z$-directed elementary dipole. Specifically, Figs. \ref{Figure5}a and \ref{Figure5}b show the numerically computed (see Appendix \ref{Sec:AppB} for details) field maps at a close distance from the metasurface, by considering the EMT model and actual conductivity modulation, respectively. Results are in fair agreement, with the differences attributable to nonlocal effects (see Sec. \ref{Sec:NL} below). In particular, they clearly display the aforementioned canalization effect, with unattenuated, diffractionless propagation of sub-wavelength features. This is markedly different from the conventional cylindrical wavefronts that would be observed in a homogeneous, isotropic case.
As a further quantitative evidence, Fig. \ref{Figure5}c shows the spatial spectrum (2-D Fourier transform) of the EMT field map, which is  essentially peaked around the theoretical EFC. Similar results, not shown for brevity, are observed for capacitive scenarios as well.

Canalization effects like those in Fig. \ref{Figure5} are of great interest for applications to high-resolution imaging.
Although the above phenomena may appear qualitatively similar to the canalization effects observed in hyperbolic metasurfaces \cite{Gomez-Diaz:2015hp,Gomez-Diaz:2015hy}, we stress that the underlying physics is completely different. While in the hyperbolic case these effects are induced by the dual character (inductive/capacitive) of the two constituents, in our case there is no contrast between the reactive parts, and the extreme anisotropy is solely induced by the gain-loss interplay.

%
\begin{figure*}
	\centering
	\includegraphics[width=\linewidth]{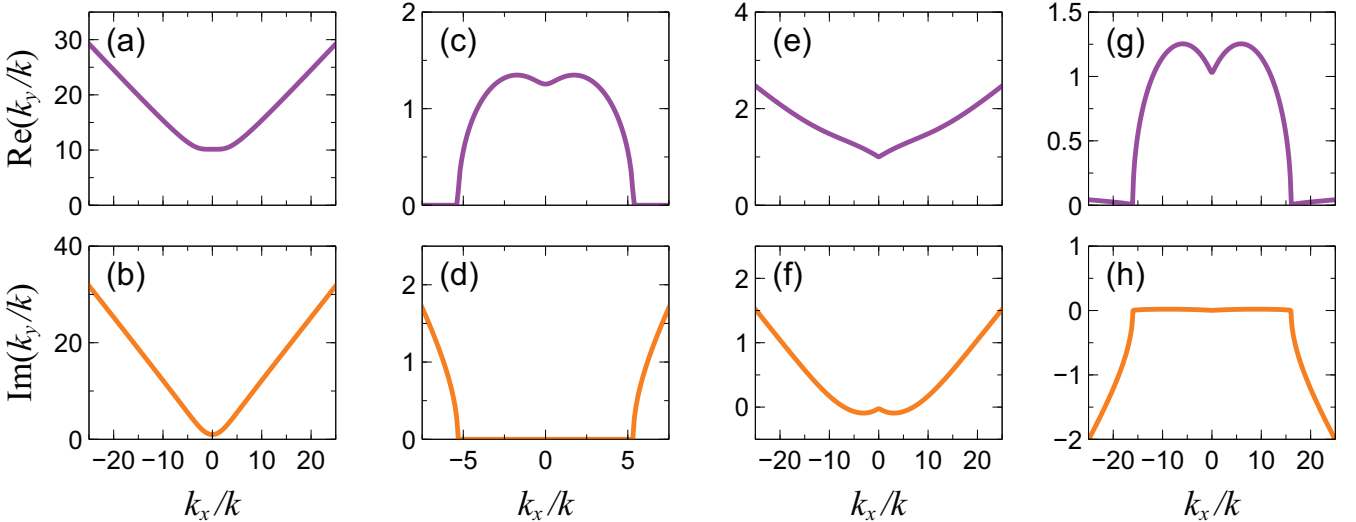}
	\caption{Representative EFCs (top panels: real parts; bottom panels: imaginary parts) for non-${\cal PT}$-symmetric configurations. 
		(a), (b) No gain: $\sigma_a\eta=1+i0.1$, $\sigma_b\eta=i0.1$, $f_a=f_b=0.5$ (${\bar \sigma}_{xx} \eta=0.5+i0.1$, ${\bar \sigma}_{yy} \eta=0.019+ i0.196$). 
		(c), (d) Gain-loss-balanced: $\sigma_a\eta=0.8+i0.608$, $\sigma_b\eta=-1+i0.1$, $f_a=0.556$, $f_b=0.444$ (${\bar \sigma}_{xx} \eta=i0.382$, ${\bar \sigma}_{yy} \eta=i2.641$). 
		(e), (f) Imperfect ${\cal PT}$ symmetry: $\sigma_a\eta=1.2+i0.1$, $\sigma_b\eta=-1+i0.1$, $f_a=f_b=0.5$ (${\bar \sigma}_{xx} \eta=0.1+i0.1$, ${\bar \sigma}_{yy} \eta=-5.95 + i6.15$). 
		(g), (h) Imperfect ${\cal PT}$ symmetry: $\sigma_a\eta=1+i0.1$, $\sigma_b\eta=-1+i0.15$, $f_a=f_b=0.5$ (${\bar \sigma}_{xx} \eta=i0.125$, ${\bar \sigma}_{yy} \eta=0.4+i8.12$).
		In view of the inherent symmetry, only the $\mbox{Re}\left(k_y\right)>0$ branches are shown. Note the different scales on the vertical axes.}
	\label{Figure6}
\end{figure*}

\subsection{Effects of Departure from  ${\cal PT}$ Symmetry}

To better understand the crucial role played by ${\cal PT}$ symmetry in establishing the extreme-anisotropy response, it is insightful to explore different configurations that do not fulfill such condition. 
Within this framework, it is also worth highlighting that the unavoidable material dispersion dictates (via causality) that the ${\cal PT}$-symmetry condition may occur only at isolated frequencies \cite{Chen:2016PT,Zyablovsky:2014ca}, and therefore the above phenomena are inherently narrowband.

\begin{figure*}
	\centering
	\includegraphics[width=\linewidth]{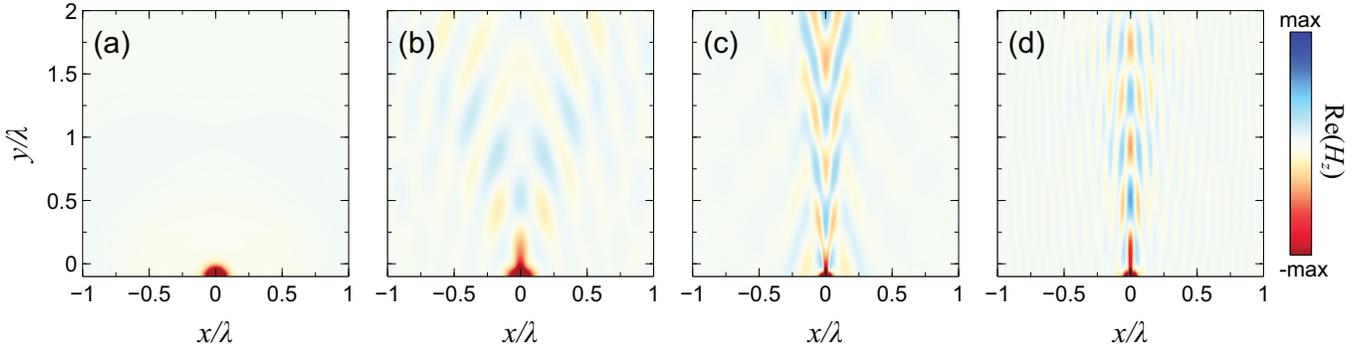}
	\caption{Numerically computed in-plane field maps [$\mbox{Re}\left(H_z\right)$], in false-color scale, pertaining to the non-${\cal PT}$-symmetric configurations in Fig. \ref{Figure6}.
		(a) No-gain, as in Figs. \ref{Figure6}a and \ref{Figure6}b. 
		(b) Gain-loss-balanced, as in Figs. \ref{Figure6}c and \ref{Figure6}d. 
		(c) Imperfect ${\cal PT}$ symmetry, as in Figs. \ref{Figure6}e and \ref{Figure6}f. 
		(d) Imperfect ${\cal PT}$ symmetry, as in Figs. \ref{Figure6}g and \ref{Figure6}h. 
		Fields are excited by a $z$-directed elementary magnetic dipole located at $x=0$, $y=-0.1\lambda$, $z=0.001\lambda$, and are computed at $z=0.01\lambda$.}
	\label{Figure7}
\end{figure*}

Figure \ref{Figure6} shows the EFCs pertaining to four representative non-${\cal PT}$-symmetric configurations of interest. Specifically, Figs. \ref{Figure6}a and \ref{Figure6}b show the (real and imaginary, respectively) results for the inductive parameter configuration in Fig. \ref{Figure3}c, but in the absence of gain ($\sigma_a\eta=1+i0.1$, $\sigma_b\eta=i0.1$). As also evident from the effective parameters (${\bar \sigma}_{xx} \eta=0.5+i0.1$, ${\bar \sigma}_{yy} \eta=0.019+ i0.196$), the extreme anisotropy is now lost and, most importantly, the propagation is severely curtailed in view of the substantial values of $\mbox{Im}\left(k_y\right)$. This is clearly visible in the corresponding field map shown in Fig. \ref{Figure7}a.
Such a stark difference from the ${\cal PT}$-symmetric case suggests interesting possibilities for dynamically reconfiguring the surface-wave response, e.g., switching between a propagating, canalized regime  and a strong damping by (de)activating the gain. 

Figures \ref{Figure6}c and \ref{Figure6}d illustrate another interesting example with parameters $\sigma_a\eta=0.8+i0.608$, $\sigma_b\eta=-1+i0.1$, $f_a=0.556$, $f_b=0.444$ satisfying the gain-loss balance conditions in (\ref{eq:lcEP}) but not the ${\cal PT}$-symmetry condition. As expected, since the effective parameters are purely reactive (inductive), the EFCs exhibit propagating and evanescent regions as in Fig. \ref{Figure3}. However, although the gain constituent is the same as in Fig. \ref{Figure3}c, the spatial bandwidth and the anisotropy are less pronounced. As can be observed in the corresponding field map shown in Fig. \ref{Figure7}b, this results in a general deterioration of the canalization effects.
Generally speaking, it can be verified numerically that, for a fixed gain constituent, among the infinite parameter configurations that fulfill the gain-loss balance conditions in (\ref{eq:lcEP}), the ${\cal PT}$-symmetry condition in (\ref{eq:PTS}) guarantees the maximum anisotropy.

As previously mentioned, when material dispersion is taken into account, the ${\cal PT}$-symmetry condition can only be perfectly satisfied at isolated frequencies. It is therefore instructive to look at the effects of moderate mismatches that can occur at close-by frequencies. For illustration, Fig. \ref{Figure6}e and \ref{Figure6}f show the EFCs for parameters as in Fig. \ref{Figure3}c when the susceptance are perfectly balanced, but there is a moderate gain-loss mismatch ($\sigma_a\eta=1.2+i0.1$, $\sigma_b\eta=-1+i0.1$). Interestingly, the resulting effective parameters (${\bar \sigma}_{xx} \eta=0.1+i0.1$, ${\bar \sigma}_{yy} \eta=-5.95 + i6.15$) exhibit simultaneously loss and gain along different directions.
Such ``indefinite`` non-Hermitian character is also observed in volumetric metamaterials \cite{Mackay:2015dc,Coppolaro:2020ep}. The  EFCs differ substantially from the perfectly  ${\cal PT}$-symmetric reference case in Fig. \ref{Figure3}c, especially in light of an imaginary part which assumes small negative values at lower wavenumbers and increasingly positive values for higher wavenumbers. Figure \ref{Figure7}c shows the corresponding field map, from which we still observe some canalization effects, though with a more visible spreading and attenuation by comparison with Fig. \ref{Figure5}.

For the same parameter configuration, Figs. \ref{Figure6}g and \ref{Figure6}h show instead the effects of the imbalance in the susceptances (while maintaining the inductive character), but  assuming now perfect gain-loss balance ($\sigma_a=1+i0.1$, $\sigma_b=-1+i0.15$). In spite of the sizable imbalance, the EFC departures from the perfectly ${\cal PT}$-symmetric reference case in Fig. \ref{Figure3}c seem less dramatic. As can be observed, there is still an extended spectral region characterized by a small positive imaginary part 
[$\mbox{Im}\left(k_y\right) \lesssim 0.02k$] and pronounced anisotropy, although the spatial bandwidth is moderately reduced. 
Quite interestingly, outside this region we observe $\mbox{Im}\left(k_y\right)<0$ which, assuming the conventional choice $\mbox{Re}\left(k_y\right)>0$ for the branch-cut, would imply amplification. However, this is not necessarily the case, as in structures mixing gain and loss an {\em a priori} choice of the branch-cut is not obvious in the presence of unbounded domains. In fact, a counterintuitive sign flip in the propagation constant has been observed at certain critical incidence conditions \cite{Sheinfux:2015ti} (see also the discussion in \cite{Coppolaro:2020ep}). In our specific case, the numerical simulations in Fig. \ref{Figure7}d indicate the presence of canalization effects accompanied by a mild attenuation, thereby suggesting that $\mbox{Re}\left(k_y\right)<0$ may be the proper choice for the branch-cut in the regions where $\mbox{Im}\left(k_y\right)<0$.

To sum up, the above examples indicate that, for moderate departures from the ${\cal PT}$-symmetry conditions, canalization effects are still attainable but with reduced resolution and propagation distance. In this respect, the gain-loss (im)balance turns out to be more critical than that in the susceptance.

%

%
\begin{figure}
	\centering
	\includegraphics[width=.8\linewidth]{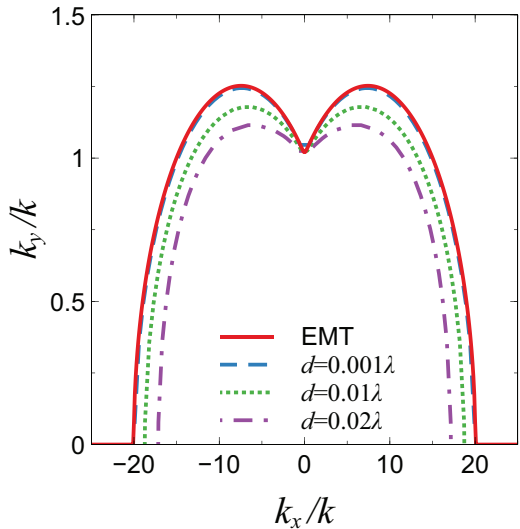}
	\caption{EFCs pertaining to a ${\cal PT}$-symmetric parameter configuration with $\sigma_a=1+i0.1$, $\sigma_b=-1+i0.1$, and $f_a=f_b=0.5$. Comparison between the EMT prediction (red-solid curve; ${\bar \sigma}_{xx}\eta=0.1$, ${\bar \sigma}_{yy}\eta=10.1$) and full-wave numerical simulations for $d=0.001\lambda$ (blue-dashed), $d=0.01\lambda$ (green-dotted), and $d=0.02\lambda$ (purple-dashed-dotted). In view of the inherent symmetry, only the $k_y>0$ branches are shown.}
	\label{Figure8}
\end{figure}

\subsection{Nonlocal Effects}
\label{Sec:NL}
As previously mentioned, the EMT approximation in (\ref{eq:EMT}) is generally accurate for deeply subwavelength modulation periods $d\ll \lambda$.
For a more quantitative assessment, Fig. \ref{Figure8} compares its predictions with the rigorous EFCs obtained via full-wave numerical simulations (see Appendix \ref{Sec:AppB}), for different values of the modulation period $d$. We observe that the results for $d=0.001\lambda$ are hardly distinguishable from the EMT prediction, whereas some visible differences progressively appear for $d=0.01\lambda$ and $d=0.02\lambda$, with changes in the local curvature and moderate reductions in the spatial bandwidth. Interestingly, the propagation constant remain purely real, thereby indicating that the ${\cal PT}$-symmetric-induced gain-loss compensation extends beyond the range of validity of the EMT approximation.

The observed departures from the EMT predictions indicate that nonlocal effects (i.e., spatial dispersion) are no longer negligible. In principle, these effects can be captured by introducing in the effective parameters some wavevector-dependent correction terms, e.g., along the lines of \cite{Correas-Serrano:2015nr}.

\section{Feasibility Issues}
\label{Sec:Implementation}
\subsection{Possible Implementation Strategies}
Although this study is essentially focused on exploring the basic phenomenology, and a practical implementation requires further investigations, 
one might wonder to what extent the parameter configurations required for ${\cal PT}$-symmetry-induced extreme anisotropy are feasible.
Within this framework, the most critical element is the gain constituent, whose implementation varies with the operational frequency of interest. For instance, at microwave frequencies, active metasurfaces typically rely on negative-resistance elements based on amplifiers \cite{Lei:2019se} or tunnel diodes \cite{Ye:2014mg}. At terahertz frequencies, an optically pumped graphene monolayer  may be a viable option, as it can support population inversion and a negative dynamic conductivity \cite{Ryzhii:2007nd}. Specifically, for typical parameters found in the literature  \cite{Chen:2016PT,Moccia:2020lw}, the real part can reach values $\sigma_g^{\prime}\approx-0.02/\eta$, while the imaginary part $\sigma_g^{\prime\prime}$ can be tuned between positive and negative values (ranging approximately from $-0.05/\eta$ to $0.05/\eta$) by acting on the frequency and quasi-Fermi energy. These figures allow in principle to attain the  large anisotropy ratios of interest here.

At optical wavelengths, gain media are typically obtained by doping host media with  organic dyes \cite{Campione:2011cm,Caligiuri:2017rs} or quantum dots \cite{Holmstrom:2010cm,Moreels:2012td,Campbell:2012tp}. To derive some basic quantitative estimates, we consider the thin-dielectric-layer model in (\ref{eq:sigd}), and invert for the complex-valued relative permittivity
\beq 
\varepsilon=\varepsilon^\prime + i\varepsilon^{\prime\prime}=1+ i\frac{\sigma \eta}{{k\Delta}}.
\eeq
Accordingly, the conditions in (\ref{eq:CAE}) for extreme anisotropy translate into
\beq
\left|1 - \varepsilon^\prime\right|k\Delta  \ll 1,\quad
\left|1 - \varepsilon^\prime\right| \ll \left|\varepsilon^{\prime\prime}\right|.
\eeq
For instance, assuming $k\Delta=0.1$, in the capacitive case, a relative permittivity $\varepsilon=1.01-i2$ would yield a normalized conductivity $\sigma\eta=-0.2-i 0.001$, like the one considered in the extreme-anisotropy case in Fig. \ref{Figure4}d. These permittivity values are in line with those attainable at infrared wavelengths by doping transparent conductive oxides (such as indium tin oxide) with lanthanides \cite{Coppolaro:2020ep,Binnemans:2009lb,Shao:2016tn,Lin:2019ir}. Similar considerations hold for the inductive case too. 

As for the  practical realizability of the required gain-loss spatial modulation, one possibility could be to rely on high-resolution selective optical pumping  
\cite{Fang:2020hs}, possibly based on digital spatial light modulators \cite{Savage:2009ds}. Alternatively, one could think of relying on a uniform optical pumping, and
patterning a thin layer of gain material with thin, lossy strips, so as to suitably overcompensate the gain in certain selected regions. 

The above considerations suggest that the gain-loss configurations of interest are within reach with current or near-future technologies, although 
 further studies are needed to develop some practical designs. In particular, 
the presence of a substrate should also be taken into account, and a simple EMT modeling may not be applicable, thereby requiring extensive numerical optimization driven by full-wave simulations. These aspects are beyond the scope of the present investigation, and will be the subject of forthcoming studies.

%
\begin{figure}
	\centering
	\includegraphics[width=.8\linewidth]{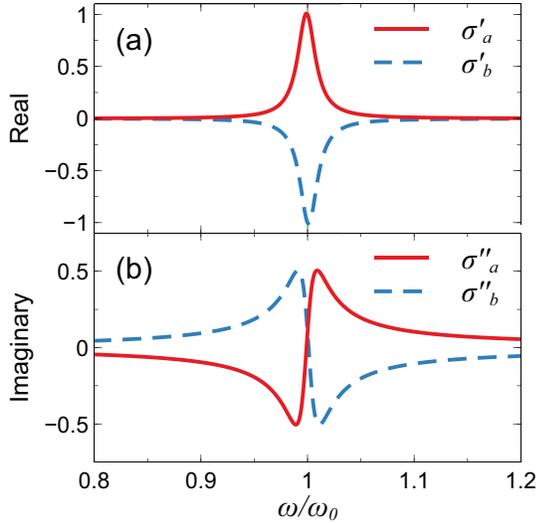}
	\caption{(a), (b) Real and imaginary parts, respectively, of dispersive models for loss and gain constituents, from (\ref{eq:displaws}). Parameters are chosen as: $A_L=A_G=2.02\cdot10^{-2}\omega_0$, $\omega_L=0.999\omega_0$, 
		$\omega_G=1.001\omega_0$, 
		$\Gamma_L=\Gamma_G=0.02\omega_0$,  in order to satisfy the conditions in Fig. \ref{Figure5}, $\sigma_a\eta=1+i0.1$, $\sigma_b\eta=-1+i0.1$, at the operational radian frequency $\omega_0$.}
	\label{Figure9}
\end{figure}

%
\begin{figure*}
	\centering
	\includegraphics[width=\linewidth]{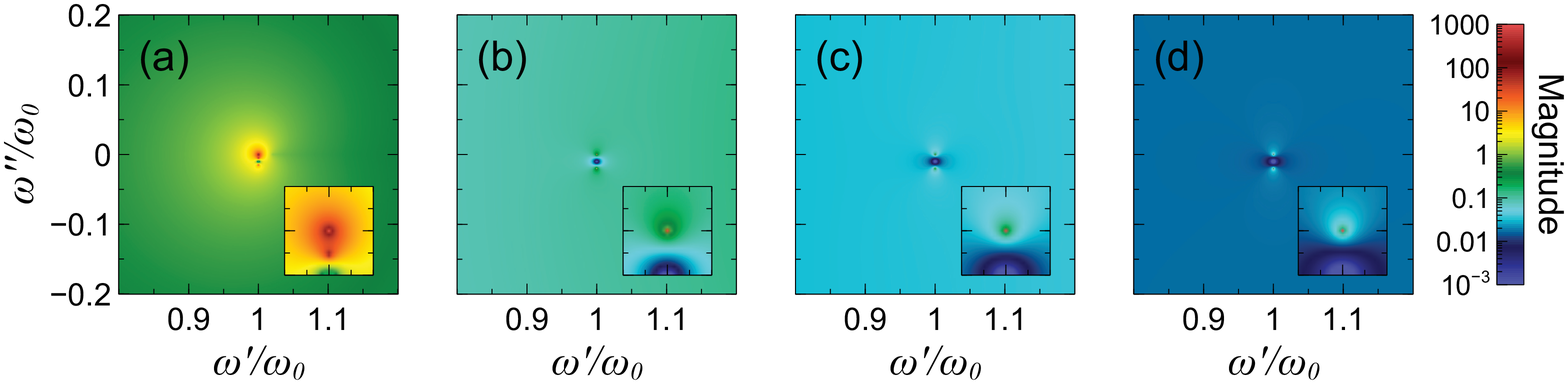}
	\caption{Representative stability maps for the parameter configuration in Fig. \ref{Figure9}.  (a), (b), (c), (d) Inverse residual (magnitude of left-hand-side reciprocal) pertaining to the dispersion equation (\ref{eq:DE}) for $k_x=0$,  $k_x=5k_0$,  $k_x=10k_0$,  $k_x=15k_0$, respectively,  in the complex ${\tilde \omega}=\omega^\prime+i\omega^{\prime\prime}$ plane; 
	$k_0=\omega_0/c$ denotes the wavenumber at the operational frequency, $k_y$ and $k_z$ are determined from the nominal EFC in Fig. \ref{Figure3}c and from  (\ref{eq:PWD}), respectively.
	The insets show magnified views around ($\omega^\prime=\omega_0$, $\omega^{\prime\prime}=0$).
		Note the presence of surface-wave poles on the real axis at $\omega^\prime=\omega_0$, and the absence of poles in the upper half-plane $\omega^{\prime\prime}>0$.}
	\label{Figure10}
\end{figure*}

\subsection{Stability}
An important issue in non-Hermitian configurations featuring gain is the potential onset of instability, manifested as self-oscillations supported by the system \cite{Zhiyenbayev:2019es}.
In previous studies dealing with planar \cite{Monticone:2016pt} and cylindrical \cite{Savoia:2017mi} non-Hermitian metasurfaces, the stability issue was addressed by introducing a physical, dispersive model for the gain constituent, and by looking at the poles of a relevant transfer function analytically continued in the complex frequency plane. Here, we follow a similar approach, and consider for the conductivities in the loss and gain regions a Lorentzian (standard and inverted, respectively) model \cite{Chen:2019ot}
\begin{subequations}
\begin{eqnarray}
\sigma_a\left(\omega\right)&=&\frac{iA_L\omega}{\eta\left(\omega^2-\omega_L^2+i\Gamma_L\omega\right)},\\
\sigma_b\left(\omega\right)&=&-\frac{iA_G\omega}{\eta\left(\omega^2-\omega_G^2+i\Gamma_G\omega\right)},
\end{eqnarray}
\label{eq:displaws}
\end{subequations}
where $A_{L,G}$, $\omega_{L,G}$ and $\Gamma_{L,G}$  denote some dimensional constants, peak radian frequencies, and damping factors, respectively. Figure \ref{Figure9} shows the above dispersion laws with parameters chosen so as to attain at an operational radian frequency $\omega_0$ the nominal values $\sigma_a\eta=1+i0.1$, $\sigma_b\eta=-1+i0.1$ considered in the canalization example of Fig. \ref{Figure5}.

We then consider the EMT dispersion equation in (\ref{eq:DEs}), which can be viewed as a pole condition for the reflection/transmission coefficient under plane-wave illumination from free space, and substitute the dispersive effective parameters computed from (\ref{eq:EP}) with (\ref{eq:displaws}). Finally, for fixed values of the wavenumbers, we look for roots of (\ref{eq:DEs}) (i.e., poles) in the complex ${\tilde \omega}=\omega^\prime+i\omega^{\prime\prime}$ plane. In view of the assumed time-harmonic convention, instability corresponds to complex poles lying in the upper half-plane $\omega^{\prime\prime}>0$.  

Figure \ref{Figure10} shows stability maps for representative values of the wavenumbers on the nominal EFC in Fig. \ref{Figure3}c. We observe the expected presence of the surface-wave pole on the real axis at the operational radian frequency $\omega^\prime=\omega_0$, and the absence of poles in the upper half-plane $\omega^{\prime\prime}>0$, which indicates that the system is unconditionally stable for any temporal excitation. The above examples only serve to illustrate that it is possible, in principle, to attain stability via suitable dispersion engineering. Clearly, different parameter choices in the dispersion models and/or the anisotropy ratios may induce the transition of some poles to the upper half-plane $\omega^{\prime\prime}>0$, thereby causing instabilities.

\section{Conclusions and Perspectives}
\label{Sec:Conclusions}
In conclusion, we have studied the surface-wave propagation for a class of non-Hermitian metasurfaces based on ${\cal PT}$-symmetric modulations of the surface conductivity. Via a simple EMT model, we have shown that  a suitably tailored gain-loss interplay can induce extreme anisotropy, giving rise to interesting canalization effects. These theoretical predictions are in good agreement with numerical full-wave simulations. Moreover, we have explored the effects of possible departures from perfect ${\cal PT}$ symmetry, as well as of nonlocality. Finally, we have addressed some preliminary feasibility issues, including the stability.

The outcomes from this study open new perspectives in the design of metasurfaces. Within this enlarged framework, the parameter space extends over the entire complex plane of the complex conductivity, and losses are not treated as second-order, detrimental effects to be minimized or compensated. Instead, their interplay with gain is harnessed to attain exotic dispersion effects, which can be dynamically controlled and/or reconfigured by acting on the gain level (e.g., via optical pumping), while maintaining a strong out-of-plane confinement. This can find a variety of potential applications in scenarios including sensing, sub-diffractive imaging and communications.

Current and future studies are aimed at exploring such applications and designing some practical implementations of the idealized configuration considered here.
Also of great interest is the study of {\em exceptional points} and lasing conditions  in these metasurfaces \cite{Feng:2017nh}.

\appendices
\section{Details on EMT Modeling}
\label{Sec:AppA}
By substituting  (\ref{eq:sigmaab}) in (\ref{eq:EMT}), we derive the real parts of the effective parameters
\begin{subequations}
\beq
{\bar \sigma}_{xx}^{\prime}=f_a \sigma_a^\prime - \left(1-f_a\right)\sigma_b^\prime, 
\label{eq:sxxp}
\eeq
\beq
{\bar \sigma}_{yy}^\prime=
\frac{f_a\sigma_a^\prime\left|\sigma_b\right|^2-\left(1-f_a\right)\sigma_b^\prime\left|\sigma_a\right|^2}{D\left(\sigma_a,\sigma_b,f_a\right)},
\label{eq:syyp}
\eeq
\end{subequations}
with
\begin{eqnarray}
D\left(\sigma_a,\sigma_b,f_a\right)&=&\left(1-f_a\right)^2\left|\sigma_a\right|^2+f_a\left|\sigma_b\right|^2\nonumber\\
&+&2f_a\left(1-f_a\right)
\left(
\sigma_a^\prime\sigma_b^\prime+\sigma_a^{\prime\prime}\sigma_b^{\prime\prime}
\right).
\label{eq:DD}
\end{eqnarray}
By recalling that, in view of the assumptions (\ref{eq:ass}), $D\left(\sigma_a,\sigma_b,f_a\right)$ in (\ref{eq:DD}) is a sum of non-negative terms, and hence never vanishes, Eqs. (\ref{eq:lcEP})--(\ref{eq:EP}) follow from zeroing (\ref{eq:sxxp}) and the numerator of (\ref{eq:syyp}).

In the ${\cal PT}$-symmetric case, the expressions in (\ref{eq:EP1}) can be directly obtained by  substituting (\ref{eq:PTSab}) in (\ref{eq:EMT}) or, equivalently, by particularizing (\ref{eq:EP}). In this last case, a limit operation is entailed in (\ref{eq:EPb}), which yields a $0/0$ indeterminate form that can be evaluated by means of the L'H\^opital's rule.

The spatial bandwidths in (\ref{eq:kx1}) and (\ref{eq:kx2}) are calculated by solving the dispersion equation (\ref{eq:DEs}) for $k_y=0$. Solving with respect to $k_z$, we find two solutions
\beq
k_{zi}=-\frac{2 k}{\eta {\bar \sigma}_{xx}},\quad
k_{zc}=-\frac{k \eta {\bar \sigma}_{yy}}{2}.
\label{eq:kzic}
\eeq
Recalling the assumed branch-cut $\mbox{Im}\left(k_z\right)\ge 0$, it is readily verified that $k_{zi}$ is the proper solution in the inductive case (${\bar \sigma}_{xx}^{\prime\prime}>0$), whereas 
$k_{zc}$ should be selected in the capacitive case (${\bar \sigma}_{yy}^{\prime\prime}<0$). The equalities in (\ref{eq:kx1}) and (\ref{eq:kx2}) follow from solving the dispersion equation  (\ref{eq:DE}) with respect to $k_x$, with the proper choice of $k_z$ in (\ref{eq:kzic}). 

\section{Details on Numerical Simulations}
\label{Sec:AppB}
The field maps in Figs. \ref{Figure5} and \ref{Figure7}, as well as the rigorous EFCs in Fig. \ref{Figure8}, are computed via numerical (finite-element) numerical simulations via the commercial software package COMSOL Multiphysics \cite{COMSOL:2015}.

For the dipole-excited configurations in Figs. \ref{Figure5} and \ref{Figure7}, we consider a 3-D computational domain of total size $3.2\lambda\times3.2\lambda\times0.35\lambda$. 
The metasurface is modeled via an impedance boundary condition at $z=0$ enforced in terms of a surface current density ($J_x=\sigma_{xx}E_x, J_y = \sigma_{yy}E_y, J_z =0$). For the configuration in Fig. \ref{Figure5}b, we model the actual conductivity modulation, with $d=0.04\lambda$ and a total 55 periods. To minimize the finite-size effects, the metasurfaces are terminated in-plane with fictitious sections of length $0.4\lambda$ with electrical conductivity tapered so as to match the free-space level. The domain is terminated (with the exception of the $z=0$ face) by perfectly matched layers of thickness $0.25\lambda$, and is discretized with an adaptive mesh, resulting into $\sim 2.3$ million degrees of freedom.
We carry out a frequency-domain analysis, by means of the the Pardiso direct solver (with default parameters). 

The spatial spectrum in Fig. \ref{Figure5}c is computed from the calculated field distribution at $z=0.01\lambda$, by means of a $2048\times512$ 2-D fast Fourier transform implemented in a Python code via the routine \texttt{fft2} available in the NumPy package \cite{NumPy:2006}.

For the EFCs in Fig. \ref{Figure8}, we consider instead a 2-D computational domain (assuming an infinite extent along the $x$-direction) including
only one period of the conductivity modulation. The structure is terminated by phase-shift walls  
along the $y$-direction, and includes a free-space layer terminated by a perfectly matched layer, both of thickness $0.5\lambda$. Once again, an adaptive mesh is applied, which yields $\sim1.7$ million degrees of freedom. In this case, we utilize the Modal Analysis, with the MUMPS direct solver and default parameters. To calculate the EFCs, we
scan the wavenumber $k_y$ over the real axis, and compute the corresponding eigenvalue $k_x$.

\ifCLASSOPTIONcaptionsoff
  \newpage
\fi




\begin{thebibliography}{10}
	\providecommand{\url}[1]{#1}
	\csname url@samestyle\endcsname
	\providecommand{\newblock}{\relax}
	\providecommand{\bibinfo}[2]{#2}
	\providecommand{\BIBentrySTDinterwordspacing}{\spaceskip=0pt\relax}
	\providecommand{\BIBentryALTinterwordstretchfactor}{4}
	\providecommand{\BIBentryALTinterwordspacing}{\spaceskip=\fontdimen2\font plus
		\BIBentryALTinterwordstretchfactor\fontdimen3\font minus
		\fontdimen4\font\relax}
	\providecommand{\BIBforeignlanguage}[2]{{%
			\expandafter\ifx\csname l@#1\endcsname\relax
			\typeout{** WARNING: IEEEtran.bst: No hyphenation pattern has been}%
			\typeout{** loaded for the language `#1'. Using the pattern for}%
			\typeout{** the default language instead.}%
			\else
			\language=\csname l@#1\endcsname
			\fi
			#2}}
	\providecommand{\BIBdecl}{\relax}
	\BIBdecl
	
	\bibitem{Yang:2019se}
	F.~Yang and Y.~Rahmat-Samii, \emph{Surface Electromagnetics: With Applications
		in Antenna, Microwave, and Optical Engineering}.\hskip 1em plus 0.5em minus
	0.4em\relax Cambridge University Press, 2019.
	
	\bibitem{Holloway:2012ao}
	C.~L. {Holloway}, E.~F. {Kuester}, J.~A. {Gordon}, J.~{O'Hara}, J.~{Booth}, and
	D.~R. {Smith}, ``An overview of the theory and applications of metasurfaces:
	The two-dimensional equivalents of metamaterials,'' \emph{IEEE Antennas
		Propagat. Mag.}, vol.~54, no.~2, pp. 10--35, Apr. 2012.
	
	\bibitem{Bao:20192d}
	Q.~Bao and H.~Hoh, \emph{{2D} {Materials} for {Photonic} and {Optoelectronic}
		{Applications}}.\hskip 1em plus 0.5em minus 0.4em\relax Elsevier Science \&
	Technology, 2019.
	
	\bibitem{Yu:2011lp}
	N.~Yu, P.~Genevet, M.~A. Kats, F.~Aieta, J.-P. Tetienne, F.~Capasso, and
	Z.~Gaburro, ``Light propagation with phase discontinuities: Generalized laws
	of reflection and refraction,'' \emph{Science}, vol. 334, no. 6054, pp.
	333--337, Oct. 2011.
	
	\bibitem{Bilow:2003gw}
	H.~J. {Bilow}, ``Guided waves on a planar tensor impedance surface,''
	\emph{IEEE Trans. Antennas Propagat.}, vol.~51, no.~10, pp. 2788--2792, Oct.
	2003.
	
	\bibitem{Patel:2013ma}
	A.~M. Patel and A.~Grbic, ``Modeling and analysis of printed-circuit tensor
	impedance surfaces,'' \emph{IEEE Trans. Antennas Propagat.}, vol.~61, no.~1,
	pp. 211--220, Jan. 2013.
	
	\bibitem{Quarfoth:2013at}
	R.~Quarfoth and D.~Sievenpiper, ``Artificial tensor impedance surface
	waveguides,'' \emph{IEEE Trans. Antennas Propagat.}, vol.~61, no.~7, pp.
	3597--3606, Jul. 2013.
	
	\bibitem{Mencagli:2015sw}
	M.~{Mencagli}, E.~{Martini}, and S.~{Maci}, ``Surface wave dispersion for
	anisotropic metasurfaces constituted by elliptical patches,'' \emph{IEEE
		Trans. Antennas Propagat.}, vol.~63, no.~7, pp. 2992--3003, Jul. 2015.
	
	\bibitem{Vakil:2011to}
	A.~Vakil and N.~Engheta, ``Transformation optics using graphene,''
	\emph{Science}, vol. 332, no. 6035, pp. 1291--1294, Jun. 2011.
	
	\bibitem{Yang:2012aa}
	R.~Yang and Y.~Hao, ``An accurate control of the surface wave using
	transformation optics,'' \emph{Opt. Express}, vol.~20, no.~9, pp. 9341--9350,
	Apr. 2012.
	
	\bibitem{Patel:2014te}
	A.~M. {Patel} and A.~{Grbic}, ``Transformation electromagnetics devices based
	on printed-circuit tensor impedance surfaces,'' \emph{IEEE Trans. Microwave
		Theory Tech.}, vol.~62, no.~5, pp. 1102--1111, May 2014.
	
	\bibitem{Mencagli:2014mb}
	M.~{Mencagli}, E.~{Martini}, D.~{González-Ovejero}, and S.~{Maci},
	``Metasurfing by transformation electromagnetics,'' \emph{IEEE Antennas
		Wireless Propagat. Lett.}, vol.~13, pp. 1767--1770, 2014.
	
	\bibitem{Martini:2015mt}
	E.~Martini, M.~Mencagli, and S.~Maci, ``Metasurface transformation for surface
	wave control,'' \emph{Philos. Trans. R. Soc. A}, vol. 373, no. 2049, p.
	20140355, Aug. 2015.
	
	\bibitem{McCall:2018ro}
	M.~McCall, J.~B. Pendry, V.~Galdi, Y.~Lai, S.~A.~R. Horsley, J.~Li, J.~Zhu,
	R.~C. Mitchell-Thomas, O.~Quevedo-Teruel, P.~Tassin, V.~Ginis, E.~Martini,
	G.~Minatti, S.~Maci, M.~Ebrahimpouri, Y.~Hao, P.~Kinsler, J.~Gratus, J.~M.
	Lukens, A.~M. Weiner, U.~Leonhardt, I.~I. Smolyaninov, V.~N. Smolyaninova,
	R.~T. Thompson, M.~Wegener, M.~Kadic, and S.~A. Cummer, ``Roadmap on
	transformation optics,'' \emph{J. Opt.}, vol.~20, no.~6, p. 063001, May 2018.
	
	\bibitem{Pendry:2006e}
	J.~B. Pendry, D.~Schurig, and D.~R. Smith, ``Controlling electromagnetic
	fields,'' \emph{Science}, vol. 312, no. 5781, pp. 1780--1782, Jun. 2006.
	
	\bibitem{Leonhardt:2006oc}
	U.~Leonhardt, ``Optical conformal mapping,'' \emph{Science}, vol. 312, no.
	5781, pp. 1777--1780, Jun. 2006.
	
	\bibitem{Yermakov:2015hw}
	O.~Y. Yermakov, A.~I. Ovcharenko, M.~Song, A.~A. Bogdanov, I.~V. Iorsh, and
	Y.~S. Kivshar, ``Hybrid waves localized at hyperbolic metasurfaces,''
	\emph{Phys. Rev. B}, vol.~91, p. 235423, Jun. 2015.
	
	\bibitem{Gomez-Diaz:2015hp}
	J.~S. Gomez-Diaz, M.~Tymchenko, and A.~Al\`u, ``Hyperbolic plasmons and
	topological transitions over uniaxial metasurfaces,'' \emph{Phys. Rev.
		Lett.}, vol. 114, p. 233901, Jun. 2015.
	
	\bibitem{Gomez-Diaz:2015hy}
	J.~S. Gomez-Diaz, M.~Tymchenko, and A.~Al\`{u}, ``Hyperbolic metasurfaces:
	surface plasmons, light-matter interactions, and physical implementation
	using graphene strips,'' \emph{Opt. Mater. Express}, vol.~5, no.~10, pp.
	2313--2329, Oct. 2015.
	
	\bibitem{Gomez-Diaz:2016fo}
	J.~S. Gomez-Diaz and A.~Al\`{u}, ``Flatland optics with hyperbolic
	metasurfaces,'' \emph{ACS Photonics}, vol.~3, no.~12, pp. 2211--2224, Dec.
	2016.
	
	\bibitem{Yang:2017hs}
	Y.~Yang, L.~Jing, L.~Shen, Z.~Wang, B.~Zheng, H.~Wang, E.~Li, N.-H. Shen,
	T.~Koschny, C.~M. Soukoulis, and H.~Chen, ``Hyperbolic spoof plasmonic
	metasurfaces,'' \emph{NPG Asia Mater.}, vol.~9, no.~8, p. e428, Aug. 2017.
	
	\bibitem{Yermakov:2018es}
	O.~Y. Yermakov, D.~V. Permyakov, F.~V. Porubaev, P.~A. Dmitriev, A.~K. Samusev,
	I.~V. Iorsh, R.~Malureanu, A.~V. Lavrinenko, and A.~A. Bogdanov, ``Effective
	surface conductivity of optical hyperbolic metasurfaces: from far-field
	characterization to surface wave analysis,'' \emph{Sci. Rep.}, vol.~8, no.~1,
	p. 14135, Sep. 2018.
	
	\bibitem{Gomez-Diaz:2017pc}
	D.~Correas-Serrano, A.~Al\`u, and J.~S. Gomez-Diaz, ``Plasmon canalization and
	tunneling over anisotropic metasurfaces,'' \emph{Phys. Rev. B}, vol.~96, p.
	075436, Aug. 2017.
	
	\bibitem{Horsley:2014od}
	S.~A.~R. Horsley and I.~R. Hooper, ``One dimensional electromagnetic waves on
	flat surfaces,'' \emph{J. Phys. D Appl. Phys.}, vol.~47, no.~43, p. 435103,
	Oct. 2014.
	
	\bibitem{Bisharat:2017ge}
	D.~J. Bisharat and D.~F. Sievenpiper, ``Guiding waves along an infinitesimal
	line between impedance surfaces,'' \emph{Phys. Rev. Lett.}, vol. 119, p.
	106802, Sep. 2017.
	
	\bibitem{Hu:2020mh}
	G.~Hu, A.~Krasnok, Y.~Mazor, C.-W. Qiu, and A.~Al\`{u}, ``Moir\'{e} hyperbolic
	metasurfaces,'' \emph{Nano Lett.}, vol.~20, no.~5, pp. 3217--3224, May 2020.
	
	\bibitem{Hu:2020tp}
	G.~Hu, Q.~Ou, G.~Si, Y.~Wu, J.~Wu, Z.~Dai, A.~Krasnok, Y.~Mazor, Q.~Zhang,
	Q.~Bao, C.-W. Qiu, and A.~Al\`u, ``Topological polaritons and photonic magic
	angles in twisted $\alpha$-{MoO} 3 bilayers,'' \emph{Nature}, vol. 582, no.
	7811, pp. 209--213, Jun. 2020.
	
	\bibitem{Bender:1998rs}
	C.~M. Bender and S.~Boettcher, ``Real spectra in non-{H}ermitian {H}amiltonians
	having {PT} symmetry,'' \emph{Phys. Rev. Lett.}, vol.~80, pp. 5243--5246,
	Jun. 1998.
	
	\bibitem{El-Ganainy:2018nh}
	R.~El-Ganainy, K.~G. Makris, M.~Khajavikhan, Z.~H. Musslimani, S.~Rotter, and
	D.~N. Christodoulides, ``Non-{Hermitian} physics and {PT} symmetry,''
	\emph{Nat. Phys.}, vol.~14, no.~1, pp. 11--19, Jan. 2018.
	
	\bibitem{Feng:2017nh}
	L.~Feng, R.~El-Ganainy, and L.~Ge, ``Non-{Hermitian} photonics based on
	parity–time symmetry,'' \emph{Nat. Photonics}, vol.~11, no.~12, pp.
	752--762, Jan. 2017.
	
	\bibitem{Fleury:2014nr}
	R.~Fleury, D.~L. Sounas, and A.~Al\`u, ``Negative refraction and planar
	focusing based on parity-time symmetric metasurfaces,'' \emph{Phys. Rev.
		Lett.}, vol. 113, p. 023903, Jul. 2014.
	
	\bibitem{Sounas:2015uc}
	D.~L. Sounas, R.~Fleury, and A.~Al\`u, ``Unidirectional cloaking based on
	metasurfaces with balanced loss and gain,'' \emph{Phys. Rev. Appl.}, vol.~4,
	p. 014005, Jul. 2015.
	
	\bibitem{Monticone:2016pt}
	F.~Monticone, C.~A. Valagiannopoulos, and A.~Al\`u, ``Parity-time symmetric
	nonlocal metasurfaces: All-angle negative refraction and volumetric
	imaging,'' \emph{Phys. Rev. X}, vol.~6, p. 041018, Oct. 2016.
	
	\bibitem{Savoia:2017mi}
	S.~Savoia, C.~A. Valagiannopoulos, F.~Monticone, G.~Castaldi, V.~Galdi, and
	A.~Al\`u, ``Magnified imaging based on non-{H}ermitian nonlocal cylindrical
	metasurfaces,'' \emph{Phys. Rev. B}, vol.~95, p. 115114, Mar. 2017.
	
	\bibitem{Chen:2016PT}
	P.-Y. Chen and J.~Jung, ``{PT} symmetry and singularity-enhanced sensing based
	on photoexcited graphene metasurfaces,'' \emph{Phys. Rev. Appl.}, vol.~5, p.
	064018, Jun. 2016.
	
	\bibitem{Sakhdari:2017pt}
	M.~Sakhdari, M.~Farhat, and P.-Y. Chen, ``{PT}-symmetric metasurfaces: wave
	manipulation and sensing using singular points,'' \emph{New J. Phys.},
	vol.~19, no.~6, p. 065002, Jun. 2017.
	
	\bibitem{Farhat:2020}
	M.~Farhat, M.~Yang, Z.~Ye, and P.-Y. Chen, ``{PT}-symmetric absorber-laser
	enables electromagnetic sensors with unprecedented sensitivity,'' \emph{ACS
		Photonics}, vol.~7, no.~8, pp. 2080--2088, Aug. 2020.
	
	\bibitem{Sakhdari:2018}
	M.~Sakhdari, N.~M. Estakhri, H.~Bagci, and P.-Y. Chen, ``Low-threshold lasing
	and coherent perfect absorption in generalized
	$\mathcal{P}\mathcal{T}$-symmetric optical structures,'' \emph{Phys. Rev.
		Appl.}, vol.~10, p. 024030, Aug. 2018.
	
	\bibitem{Moccia:2020lw}
	M.~Moccia, G.~Castaldi, A.~Al\`{u}, and V.~Galdi, ``Line waves in
	non-{H}ermitian metasurfaces,'' \emph{ACS Photonics}, vol.~7, no.~8, pp.
	2064--2072, Aug. 2020.
	
	\bibitem{Kolkowski:2020lr}
	R.~{Kolkowski} and A.~F. {Koenderink}, ``Lattice resonances in optical
	metasurfaces with gain and loss,'' \emph{P. IEEE}, vol. 108, no.~5, pp.
	795--818, May 2020.
	
	\bibitem{Coppolaro:2020ep}
	M.~Coppolaro, M.~Moccia, V.~Caligiuri, G.~Castaldi, N.~Engheta, and V.~Galdi,
	``Extreme-parameter non-{H}ermitian dielectric metamaterials,'' \emph{ACS
		Photonics}, vol.~7, no.~9, pp. 2578--2588, Sep. 2020.
	
	\bibitem{Bisharat:2018ml}
	D.~J. Bisharat and D.~F. Sievenpiper, ``Manipulating line waves in flat
	graphene for agile terahertz applications,'' \emph{Nanophotonics}, vol.~7,
	no.~5, pp. 893--903, May 2018.
	
	\bibitem{Mattheakis:2016en}
	M.~Mattheakis, C.~A. Valagiannopoulos, and E.~Kaxiras, ``Epsilon-near-zero
	behavior from plasmonic dirac point: Theory and realization using
	two-dimensional materials,'' \emph{Phys. Rev. B}, vol.~94, p. 201404, Nov.
	2016.
	
	\bibitem{Forati:2014ph}
	E.~Forati, G.~W. Hanson, A.~B. Yakovlev, and A.~Al\`u, ``Planar hyperlens based
	on a modulated graphene monolayer,'' \emph{Phys. Rev. B}, vol.~89, p. 081410,
	Feb. 2014.
	
	\bibitem{Sihvola:1999em}
	A.~H. Sihvola, \emph{Electromagnetic {Mixing} {Formulas} and
		{Applications}}.\hskip 1em plus 0.5em minus 0.4em\relax IEE, 1999.
	
	\bibitem{Ortiz:2013sc}
	J.~D. {Ortiz}, J.~D. {Baena}, V.~{Losada}, F.~{Medina}, R.~{Marqu\'{e}s}, and
	J.~L.~A. {Quijano}, ``Self-complementary metasurface for designing narrow
	band pass/stop filters,'' \emph{IEEE Microw. Wirel. Compon. Lett.}, vol.~23,
	no.~6, pp. 291--293, Jun. 2013.
	
	\bibitem{Gonzalez:2015sw}
	D.~{Gonz\'{a}lez-Ovejero}, E.~{Martini}, and S.~{Maci}, ``Surface waves
	supported by metasurfaces with self-complementary geometries,'' \emph{IEEE
		Trans. Antennas Propagat.}, vol.~63, no.~1, pp. 250--260, Jan. 2015.
	
	\bibitem{Baena:2017ba}
	J.~D. {Baena}, S.~B. {Glybovski}, J.~P. {del Risco}, A.~P. {Slobozhanyuk}, and
	P.~A. {Belov}, ``Broadband and thin linear-to-circular polarizers based on
	self-complementary zigzag metasurfaces,'' \emph{IEEE Trans. Antennas
		Propagat.}, vol.~65, no.~8, pp. 4124--4133, Aug. 2017.
	
	\bibitem{Jiang:2015ev}
	H.~Jiang, W.~Liu, K.~Yu, K.~Fang, Y.~Sun, Y.~Li, and H.~Chen, ``Experimental
	verification of loss-induced field enhancement and collimation in anisotropic
	$\ensuremath{\mu}$-near-zero metamaterials,'' \emph{Phys. Rev. B}, vol.~91,
	p. 045302, Jan. 2015.
	
	\bibitem{Zyablovsky:2014ca}
	A.~A. Zyablovsky, A.~P. Vinogradov, A.~V. Dorofeenko, A.~A. Pukhov, and A.~A.
	Lisyansky, ``Causality and phase transitions in $\mathcal{PT}$-symmetric
	optical systems,'' \emph{Phys. Rev. A}, vol.~89, p. 033808, Mar. 2014.
	
	\bibitem{Mackay:2015dc}
	T.~G. Mackay and A.~Lakhtakia, ``Dynamically controllable anisotropic
	metamaterials with simultaneous attenuation and amplification,'' \emph{Phys.
		Rev. A}, vol.~92, p. 053847, Nov. 2015.
	
	\bibitem{Sheinfux:2015ti}
	H.~Herzig~Sheinfux, B.~Zhen, I.~Kaminer, and M.~Segev, ``Total internal
	reflection in gain media,'' in \emph{CLEO: 2015}.\hskip 1em plus 0.5em minus
	0.4em\relax Optical Society of America, 2015, p. FM2D.3.
	
	\bibitem{Correas-Serrano:2015nr}
	D.~Correas-Serrano, J.~S. Gomez-Diaz, M.~Tymchenko, and A.~Al\`{u}, ``Nonlocal
	response of hyperbolic metasurfaces,'' \emph{Opt. Express}, vol.~23, no.~23,
	pp. 29\,434--29\,448, Nov. 2015.
	
	\bibitem{Lei:2019se}
	L.~Chen, Q.~Ma, H.~B. Jing, H.~Y. Cui, Y.~Liu, and T.~J. Cui, ``Space-energy
	digital-coding metasurface based on an active amplifier,'' \emph{Phys. Rev.
		Applied}, vol.~11, p. 054051, May 2019.
	
	\bibitem{Ye:2014mg}
	D.~Ye, K.~Chang, L.~Ran, and H.~Xin, ``\BIBforeignlanguage{en}{Microwave gain
		medium with negative refractive index},'' \emph{\BIBforeignlanguage{en}{Nat.
			Commun.}}, vol.~5, p. 5841, Dec. 2014.
	
	\bibitem{Ryzhii:2007nd}
	V.~Ryzhii, M.~Ryzhii, and T.~Otsuji, ``Negative dynamic conductivity of
	graphene with optical pumping,'' \emph{J. Appl. Phys.}, vol. 101, no.~8, p.
	083114, Apr. 2007.
	
	\bibitem{Campione:2011cm}
	S.~Campione, M.~Albani, and F.~Capolino, ``Complex modes and near-zero
	permittivity in {3D} arrays of plasmonic nanoshells: loss compensation using
	gain,'' \emph{Opt. Mater. Express}, vol.~1, no.~6, pp. 1077--1089, Oct. 2011.
	
	\bibitem{Caligiuri:2017rs}
	V.~Caligiuri, L.~Pezzi, A.~Veltri, and A.~De~Luca, ``Resonant gain
	singularities in {1D} and {3D} metal/dielectric multilayered
	nanostructures,'' \emph{ACS Nano}, vol.~11, no.~1, pp. 1012--1025, Jan. 2017.
	
	\bibitem{Holmstrom:2010cm}
	P.~Holmstr\"om, L.~Thyl\'{e}n, and A.~Bratkovsky, ``Composite metal/quantum-dot
	nanoparticle-array waveguides with compensated loss,'' \emph{Appl. Phys.
		Lett.}, vol.~97, no.~7, p. 073110, Aug. 2010.
	
	\bibitem{Moreels:2012td}
	I.~Moreels, D.~Kruschke, P.~Glas, and J.~W. Tomm, ``The dielectric function of
	{PbS} quantum dots in a glass matrix,'' \emph{Opt. Mater. Express}, vol.~2,
	no.~5, pp. 496--500, May 2012.
	
	\bibitem{Campbell:2012tp}
	S.~D. Campbell and R.~W. Ziolkowski, ``The performance of active coated
	nanoparticles based on quantum-dot gain media,'' \emph{Adv. Optoelectron.},
	vol. 2012, pp. 1--6, Jan. 2012.
	
	\bibitem{Binnemans:2009lb}
	K.~Binnemans, ``Lanthanide-based luminescent hybrid materials,'' \emph{Chem.
		Rev.}, vol. 109, no.~9, pp. 4283--4374, Aug. 2009.
	
	\bibitem{Shao:2016tn}
	W.~Shao, G.~Chen, A.~Kuzmin, H.~L. Kutscher, A.~Pliss, T.~Y. Ohulchanskyy, and
	P.~N. Prasad, ``Tunable narrow band emissions from dye-sensitized
	core/shell/shell nanocrystals in the second near-infrared biological
	window,'' \emph{J. Am. Chem. Soc.}, vol. 138, no.~50, pp. 16\,192--16\,195,
	Dec. 2016.
	
	\bibitem{Lin:2019ir}
	H.~Lin, D.~Xu, Y.~Li, L.~Yao, L.~Xu, Y.~Ma, S.~Yang, and Y.~Zhang, ``Intense
	red upconversion luminescence in {E}r3+-sensitized particles through
	confiniing the 1532 nm excitation energy,'' \emph{J. Lumin.}, vol. 216, p.
	116731, Dec. 2019.
	
	\bibitem{Fang:2020hs}
	X.~Fang, K.~Wei, T.~Zhao, Y.~Zhai, D.~Ma, B.~Xing, Y.~Liu, and Z.~Xiao, ``High
	spatial resolution multi-channel optically pumped atomic magnetometer based
	on a spatial light modulator,'' \emph{Opt. Express}, vol.~28, no.~18, pp.
	26\,447--26\,460, Aug. 2020.
	
	\bibitem{Savage:2009ds}
	N.~Savage, ``Digital spatial light modulators,'' \emph{Nat. Photonics}, vol.~3,
	no.~3, pp. 170--172, Mar. 2009.
	
	\bibitem{Zhiyenbayev:2019es}
	Y.~Zhiyenbayev, Y.~Kominis, C.~Valagiannopoulos, V.~Kovanis, and A.~Bountis,
	``Enhanced stability, bistability, and exceptional points in saturable active
	photonic couplers,'' \emph{Phys. Rev. A}, vol. 100, p. 043834, Oct. 2019.
	
	\bibitem{Chen:2019ot}
	S.~Chen, P.~K\"uhne, V.~Stanishev, S.~Knight, R.~Brooke, I.~Petsagkourakis,
	X.~Crispin, M.~Schubert, V.~Darakchieva, and M.~P. Jonsson, ``On the
	anomalous optical conductivity dispersion of electrically conducting
	polymers: ultra-wide spectral range ellipsometry combined with a
	{D}rude--{L}orentz model,'' \emph{J. Mater. Chem. C}, vol.~7, pp. 4350--4362,
	Apr. 2019.
	
	\bibitem{COMSOL:2015}
	{COMSOL Group}, \emph{COMSOL Multiphysics: Version 5.1}.\hskip 1em plus 0.5em
	minus 0.4em\relax COMSOL, Stockholm, 2015.
	
	\bibitem{NumPy:2006}
	\BIBentryALTinterwordspacing
	T.~Oliphant, ``{NumPy}: A guide to {NumPy},'' USA: Trelgol Publishing, 2006.
	[Online]. Available: \url{http://www.numpy.org/}
	\BIBentrySTDinterwordspacing
	
\end{thebibliography}


\end{document}